\documentclass[useAMS,usenatbib]{mn2e}

\usepackage{amsmath,array,amsfonts}
\usepackage{amssymb}
\usepackage{graphicx}
\usepackage{natbib}
\usepackage{pstricks}
\usepackage{verbatim}
\usepackage{multirow}
\usepackage{bbold}

\newcommand{\ltsima} {$\; \buildrel < \over \sim \;$}
\newcommand{\gtsima} {$\; \buildrel > \over \sim \;$}
\newcommand{\lta} {\lower.5ex\hbox{\ltsima}}
\newcommand{\gta} {\lower.5ex\hbox{\gtsima}}

\title[Dipole modulation in LSS]{Searching for a dipole modulation in the large-scale structure of the Universe}

\author[R. Fern\'{a}ndez-Cobos et al.]{R. Fern\'{a}ndez-Cobos$^1$$^,$$^2$\thanks{e-mail:cobos@ifca.unican.es}, P. Vielva$^1$, D. Pietrobon$^3$,
                                       A. Balbi$^4$$^,$$^5$, \newauthor E. Mart\'inez-Gonz\'alez$^1$, R. B. Barreiro$^1$  \\
$^1$     Instituto de F\'isica de Cantabria, CSIC-Universidad de Cantabria, Avda. de los Castros s/n, 39005 Santander, Spain.\\
$^2$     Dpto. de F\'isica Moderna, Universidad de Cantabria, Avda. los Castros s/n, 39005 Santander, Spain.\\
$^3$     Jet Propulsion Laboratory, California Institute of Technology, Pasadena, CA 91109, USA.\\
$^4$     Dipartimento di Fisica, Universit\`a di Roma ``Tor Vergata'', Via Della Ricerca Scientifica, 00133 Roma, Italy.\\
$^5$     INFN, Sezione di Roma Tor Vergata, 00133 Roma, Italy.}

\date{Accepted  Received ; in original form }

\begin{document}

\maketitle

\begin{abstract}
Several statistical anomalies in the CMB temperature anisotropies seem to defy the assumption of a homogeneous and isotropic universe. In particular, 
a dipole modulation has been detected both in \textit{WMAP} and \textit{Planck} data. We adapt the
methodology proposed by Eriksen et al. (2007) on CMB data to galaxy surveys, tracing the large-scale structure. We analyse the NRAO 
VLA Sky Survey (NVSS) data at a resolution of $\sim 2^{\circ}$ for three different flux thresholds: 
$2.5$, $5.0$ and $10.0$ mJy respectively. No evidence of a dipole modulation is found. This result suggests that the origin of 
the dipole asymmetry found in the CMB cannot be assigned to secondary anisotropies produced at redshifts around $z = 1$. However, it could still have been generated 
at redshifts higher or lower, such as the integrated Sachs-Wolfe effect produced by the local structures. Other all-sky surveys, like the infrared \textit{WISE} catalogue,
 could help to explore with a high sensitivity a redshift interval closer than the one probed with NVSS.
\end{abstract}
\begin{keywords}
methods: data analysis - cosmic microwave background
\end{keywords}
\section{Introduction}
\label{sec:Introduction}
The acceptance of the cosmological principle requires that the Universe, on sufficiently large scales, 
is statistically isotropic and homogeneous. This assumption is widely supported by both the results 
obtained by the ESA \textit{Planck} satellite \citep{PlanckXV2013} and the galaxy catalogues generated 
by the large-covering sky surveys carried out during the last decade, such as the Sloan Digital Sky Survey \citep[SDSS;][]{York2000}. 
However, several anomalies that seem to deviate from the orthodoxy were detected in the \textit{WMAP} data, 
and recently confirmed by \textit{Planck}  \citep{PlanckXXIII2013}. 
These hints of anomalous behaviour in the cosmic microwave background (CMB) observations include
 the detection of a non-Gaussian cold spot in the Southern galactic hemisphere \citep[e.g.,][]{Vielva2004, Cruz2005},
 an alignment between the quadrupole and the octopole \citep[e.g.,][]{Oliveira2004, Schwarz2004, Land2005}, 
the lack of large-scale power \citep[e.g.,][]{Berera1998, Spergel2003, Copi2007} or the so-called hemispherical asymmetry \citep[e.g.,][]{Eriksen2004, Hansen2004, Park2004}.

In this paper, we focus on the characterisation of this North-South hemispherical asymmetry. In particular, this anomaly was discovered
 in the first-year \textit{WMAP} data, in which several estimators calculated on different regions 
of the sphere show an anisotropic behaviour (see the previous references). It seems that the CMB anisotropy pattern in the Northern hemisphere has a lack of structure with 
respect to the Southern hemisphere. \citet{Inoue2006} explored the possibility that the CMB anomalies could be caused by the presence 
of local voids. In particular, they showed that the North-South anisotropy is compatible with an asymmetric distribution of these 
local voids between the two hemispheres. In a later paper, \citet{Hansen2009} extended the analysis of \citet{Eriksen2004} and \citet{Hansen2004}
to higher angular multipoles, and confirmed this asymmetry. More recently, the \textit{Planck} collaboration \citep{PlanckXXIII2013} repeated the analysis 
of \citet{Eriksen2005}, by computing the two-, three- and four-correlation functions of the data, and they obtained results in agreement with those provided with \textit{WMAP} 
data. These hemispherical differences have been
supported by subsequent analysis \citep[e.g.,][]{Santos2013, Akrami2014}.

Moreover, \citet{Gordon2005} proposed an effective model in terms of a dipole modulation that would characterise the hypothetical case in which the statistical 
isotropy is spontaneously broken in the fluctuations observed from a given position. \citet{Bennett2011} explicity distinguished between a hemispherical asymmetry,
in which the power spectrum changes discontinuously across a large circle on the sky, and a dipole asymmetry, in which the fluctuations are smoothly modulated
by a cosine function across the sky. Assuming the \citet{Gordon2005} parameterization, \citet{Eriksen2007} estimated the amplitude and 
the direction of the modulation in the 3-year \textit{WMAP} data. They found evidence of a preferred direction that in galactic coordinates points towards 
$(l,b) = (225^{\circ},-27^{\circ})$. In a later paper, \citet{Hoftuft2009} concluded that the best-fit modulation amplitude for $\ell \leq 64$ and the ILC of 
the 5-year \textit{WMAP} data with the KQ85 mask is $A_m=0.072\pm 0.022$ and the preferred direction points towards 
$(l,b) = (224^{\circ},-22^{\circ}) \pm 24^{\circ}$. Finally, the \textit{Planck} collaboration, also following the methodology presented by \citet{Eriksen2007}, detected 
this dipole modulation with consistent results with those obtained with the \textit{WMAP} data \citep{PlanckXXIII2013}. Indeed, the Bipolar Spherical Harmonic
 (BipoSH) formalism has also been applied by \citet{PlanckXXIII2013} to
demonstrate that a dipole modulation term is sufficient to account for this asymmetry. 

Several  theoretical proposals that try to explain this dipole asymmetry, such as models with a modulation of the reionization optical depth or a modulated
scale-dependent isocurvature component, can also be found in the literature \citep[see, e.g.,][and references therein]{Dai2013}. For instance, \citet{Ackerman2007} studied the
possibility of an asymmetric pattern in the CMB as an imprint of a primordial preferred direction during inflation. A theoretical model developed within a framework
of single-field inflation was proposed by \citet{Donoghue2009}. \citet{McDonald2013} presented a modulated reheating model. An alternative model in which an spectator field has a fast roll phase
was presented by \citet{Mazumdar2013} and an inflation version with a contracting phase was proposed by \citet{Liu2013}. \citet{Damico2013} analysed a model of inflation in which the particle production leads to a dipole 
modulation. And, more recently, \citet{Liddle2013} proposed a marginally-open model for the universe which could explain, in particular, the dipole power asymmetry.

The main motivation of this analysis is to shed light on the possible cosmological origin of the CMB dipole asymmetry 
by checking whether it is also detected in the large-scale structure (LSS) data. The galaxy distribution is an exceptional observable to collate a possible isotropy
breaking, because it traces the gravitational potencial \citep[see, for instance,][]{Abolhasani2014}.
In the case that some catalogues were analysed and no preferred direction was found, the CMB asymmetry could be due to, for instance, a secondary anisotropy located into a different redshift 
range than those which were considered. In fact, several authors addressed this problem by searching in the integrated Sachs-Wolfe field the cause of the preferred axis
found in the CMB \citep[see, e.g.,][]{Francis2010, Rassat2013}. 

In particular, we analyse here the galaxy distribution provided by the NRAO VLA Sky Survey \footnote[3]{http://www.cv.nrao.edu/nvss/} \citep[NVSS;][]{Condon1998}. 
This survey is appropriate to carry out cosmological analysis, such as cross-correlations with CMB data to characterise the
integrated Sachs-Wolfe effect \citep[e.g.,][]{Boughn2004, Vielva2006, McEwen2007, Schiavon2012, Giannantonio2012, Barreiro2013, PlanckXIX2013} or to find constrains on 
primordial non-Gaussianity \citep[see, for instance,][]{Xia2011, Giannantonio2014, Marcos2013}.

This paper is structured as follows. The methodology is explained in Section \ref{sec:Method}. Then, we describe the NVSS data in section \ref{sec:data}.
The results obtained both for simulated and real data are presented in Section \ref{sec:results_dip}. Finally, conclusions are shown in Section \ref{sec:Conclusions}.
    
%
\section{The method}
\label{sec:Method}
The Bayesian approaches computed with Monte Carlo (MCMC) samplers are widely used in the CMB data analysis
 \citep[e.g.,][]{Jewell2004, Wandelt2004, Eriksen2004c, Chu2005}. We employ here an adapted version of the 
framework used by \citet{Eriksen2007} to characterise the dipole asymmetry in LSS data.

A phenomenological model, like the one proposed by \citet{Gordon2005} for the CMB data, is assumed to describe the fluctuations 
of the number density of galaxies $\boldsymbol{\delta}$ in terms of a modulation of the isotropic signal in the direction $\mathbf{x_m}$: 
\begin{equation}
\delta(\mathbf{x}) = d_k(\mathbf{x},\mathbf{x_k}) + \left[ 1+d_m(\mathbf{x},\mathbf{x_m}) \right] \left[s(\mathbf{x}) + n_p(\mathbf{x}) \right], 
\end{equation}
where $\delta(\mathbf{x}) \equiv \left[ n(\mathbf{x}) - \bar{n} \right]/\bar{n}$, being $n(\mathbf{x})$ the number of counts integrated 
in the area of the pixel centred in the direction $\mathbf{x}$ and $\bar{n}$ the average number of counts per pixel. 
The isotropic Gaussian random field of the number density fluctuations predicted by the standard model is represented by $s(\mathbf{x})$, 
while $n_p(\mathbf{x})$ refers to the intrinsic Poisson noise of the galaxy count. The dipole terms are considered as 
$d_i(\mathbf{x_i},\mathbf{x}) \equiv A_i
\cos{\theta_{\mathbf{x},\mathbf{x_i}}}$, with
$\theta_{\mathbf{x},\mathbf{x_i}}$ denoting the angular distance 
between the unitary vectors $\mathbf{x}$ and $\mathbf{x_i}$ and $i = \left\{ k,m \right\}$. The subscript $m$ refers to a dipole
modulation with a preferred direction $\mathbf{x_m}$, and the subscript $k$ denotes the additive dipole term in the direction $\mathbf{x_k}$.
This latter term accounts for both the dominant kinetical component because of our local motion and the residual dipole due to a partial-sky coverage. 
As the well known CMB dipole anisotropy, a corresponding effect is also present in the number density of galaxies due to a combination of Doppler boosting and relativistic aberration 
\citep[see, for instance,][]{Singal2011, Gibelyou2012, Rubart2013, Tiwari2013}. 

The model depends on $6$ parameters which determine the dimensionless amplitudes of the dipole terms $A_i$ and their orientations 
$\mathbf{x_i}$, which are parameterized by two angles. Contrary to what happens in the CMB experiments, 
where the instrumental noise is an extra contribution from the detector electronics, the noise term here represents 
an uncertainty which is intrinsic to the measure. For this reason, the main difference with respect to the model used 
by \citet{Hoftuft2009} is that we also modulate this noise term. The average number of counts per pixel is large enough 
to consider that the noise contribution is well described by a Gaussian distribution. The dimensionless noise term is assumed
to have variance $\sigma^2_p = 1/\bar{n}$. 

The previous definition of the data set implies that the modulated fluctuations of the number density of galaxies 
$\boldsymbol{\delta}$ are well characterised by an anisotropic, but still Gaussian, random field. The 
covariance matrix of these data can be expressed as 
\begin{equation}
\mathrm{C}_{\mathrm{mod}}(\mathbf{x_{\alpha}},\mathbf{x_{\beta}}) = f_{\mathbf{x_{\alpha}}} \left[ \mathrm{C}_{\mathrm{iso}}(\mathbf{x_{\alpha}},\mathbf{x_{\beta}}) + \mathrm{N}(\mathbf{x_{\alpha}},\mathbf{x_{\beta}}) \right] f_{\mathbf{x_{\beta}}},   
\end{equation}   
where $f_{\mathbf{x_j}}$ depends on the dipole modulation parameters, $f_{\mathbf{x_j}} \equiv 1+A_m \cos{\theta_{\mathbf{x_j},\mathbf{x_m}}}$ (with $j = \alpha$ or $\beta$), and $\mathrm{\mathbf{N}}$ is a diagonal matrix that takes into account the Poisson noise where all nonzero terms are equal to $\sigma^2_p$. 

The covariance matrix of the isotropic signal $\mathrm{\mathbf{C}}_{\mathrm{iso}}$ is constructed as 
\begin{equation}
\label{eq:cov_iso}
\mathrm{C}_{\mathrm{iso}}(\mathbf{x_{\alpha}},\mathbf{x_{\beta}}) = \dfrac{1}{4\pi} \sum_{\ell=2}^{\ell_{\mathrm{max}}} {\left( 2\ell+1 \right)C_{\ell}^{\mathrm{GG}}P_{\ell}(\cos{\theta_{\mathbf{x_{\alpha}},\mathbf{x_{\beta}}}})}, 
\end{equation}
where $C_{\ell}^{\mathrm{GG}}$ is a theoretical model for the power spectrum of the data and $P_{\ell}$ denotes the Legendre polynomial of order $\ell$. The noise term is large enough  to solve the 
regularization problems mentioned by \citet{Eriksen2007}, so we do not need to add any artificial contribution. Applying a $\chi^2$ test to ensure the consistency between the covariance matrix of the isotropic 
signal and the simulated maps, it was found that the optimum value for $\ell_{\mathrm{max}}$ is $3 \mathrm{N_{side}} -1$, where $ \mathrm{N_{side}}$ is an integer defined in the  \textsc{HEALPix} tessellation 
\citep{Gorski2005} so that the number of pixels needed to cover the sphere is $\mathrm{N_{pix}}=12\mathrm{N^2_{side}}$.

The parameters of the model are estimated by maximizing the posterior probability. Since both the number density of galaxies 
and its intrinsic noise are assumed to be Gaussian, the log-likelihood can be written as
\begin{equation}
- \log{\mathfrak{L}} = \dfrac{1}{2}\left( \mathbf{d}^T\mathrm{\mathbf{C}}_{\mathrm{mod}}^{-1}\mathbf{d} + \log{|\mathbf{d}|}\right),
\label{eq:likelihood}
\end{equation} 
ignoring an irrelevant constant. In our particular case, both the covariance matrix $\mathrm{\mathbf{C}}_{\mathrm{mod}}$ and the data set 
$\mathbf{d} \equiv \boldsymbol{\delta} - \mathbf{d_k}$ depend on the parameters of the modelling. To estimate the amplitude and the orientation 
of the dipole modulation, we marginalise over the other parameters, including the three ones which characterise the amplitude and direction of
 the additive dipole.   

We use an adapted version of the \textsc{CosmoMC} \footnote[4]{http://cosmologist.info/cosmomc/} code \citep{Lewis2002} to sample the likelihood
described by the Equation \ref{eq:likelihood}. In practice, the covariance matrix $\mathrm{\mathbf{C}}_{\mathrm{mod}}$ depends on three parameters 
 ($A_m$, $\cos{\theta}$, $\phi$), where $\theta$ is the colatitude and $\phi$ the galactic longitude. Actually, the solution 
is degenerate because of the symmetry of the problem. In case of detection, there would be a solution with positive amplitude and 
a certain orientation, and its complementary, which corresponds to a rotation of $\pi$ in the direction and a sign change in the amplitude. To avoid this 
inconvenience, we constrain the galactic longitude into a range of size $\pi$ to select only a half of the sphere. This $\pi$ range for the galactic longitude
is determined after exploring the whole $2\pi$ range with some preliminary runs. The final range is selected such as the maximum value of the posterior of this parameter
is approximately centred on it.

\section{The NVSS data}
\label{sec:data}
\begin{figure}
\centering
\includegraphics[scale=0.3]{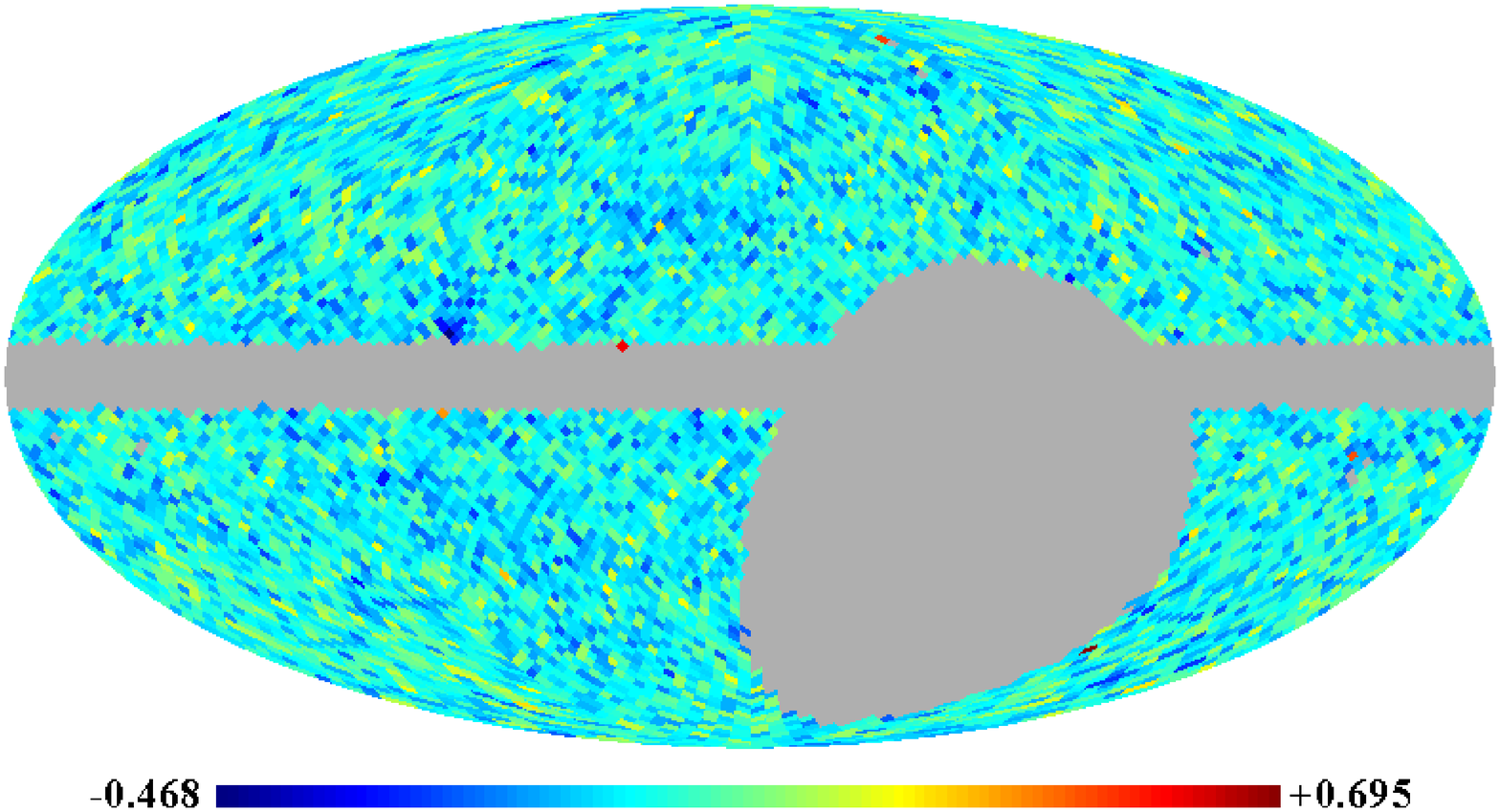}  \\

\includegraphics[scale=0.3]{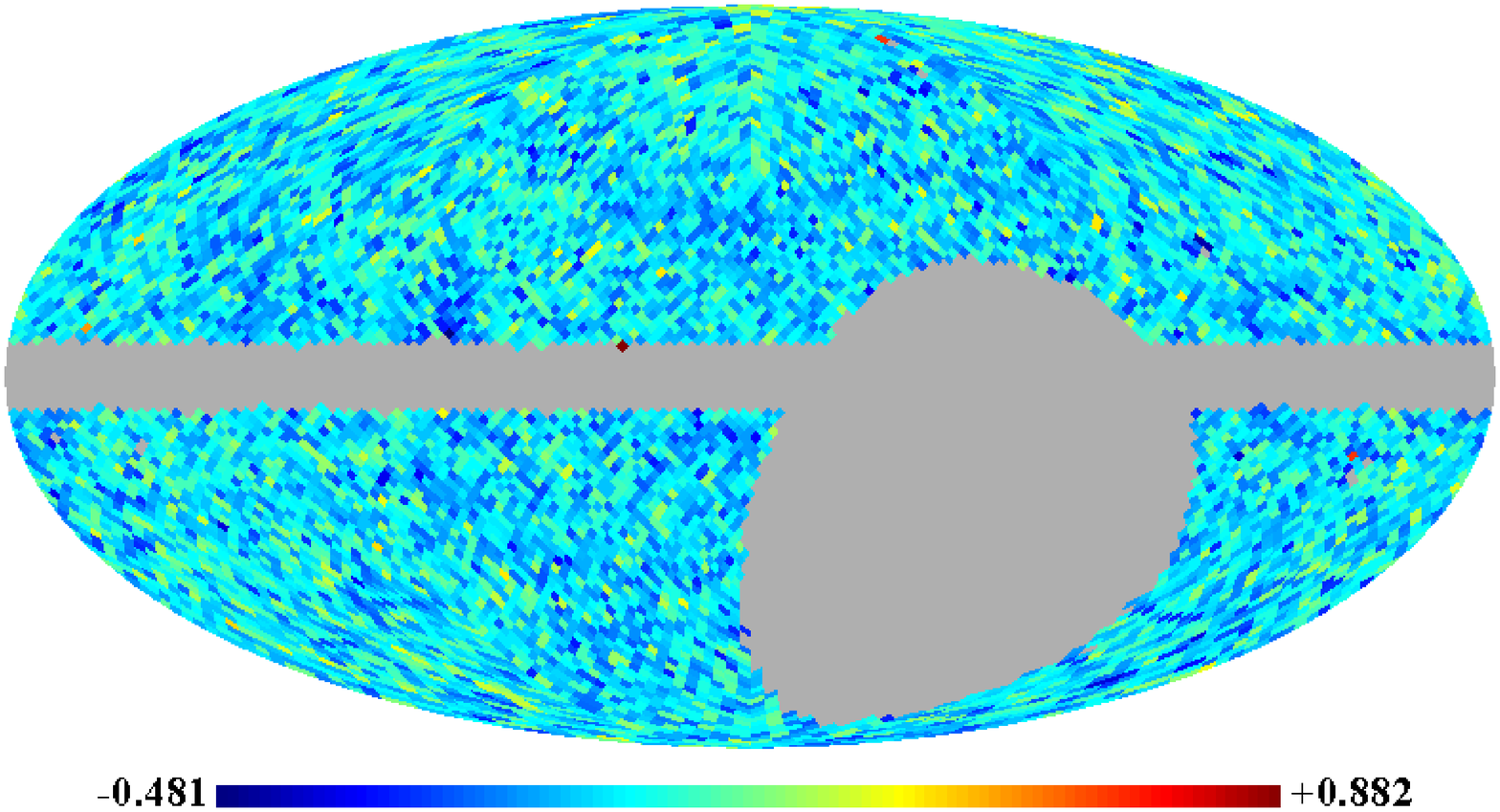}  \\

\includegraphics[scale=0.3]{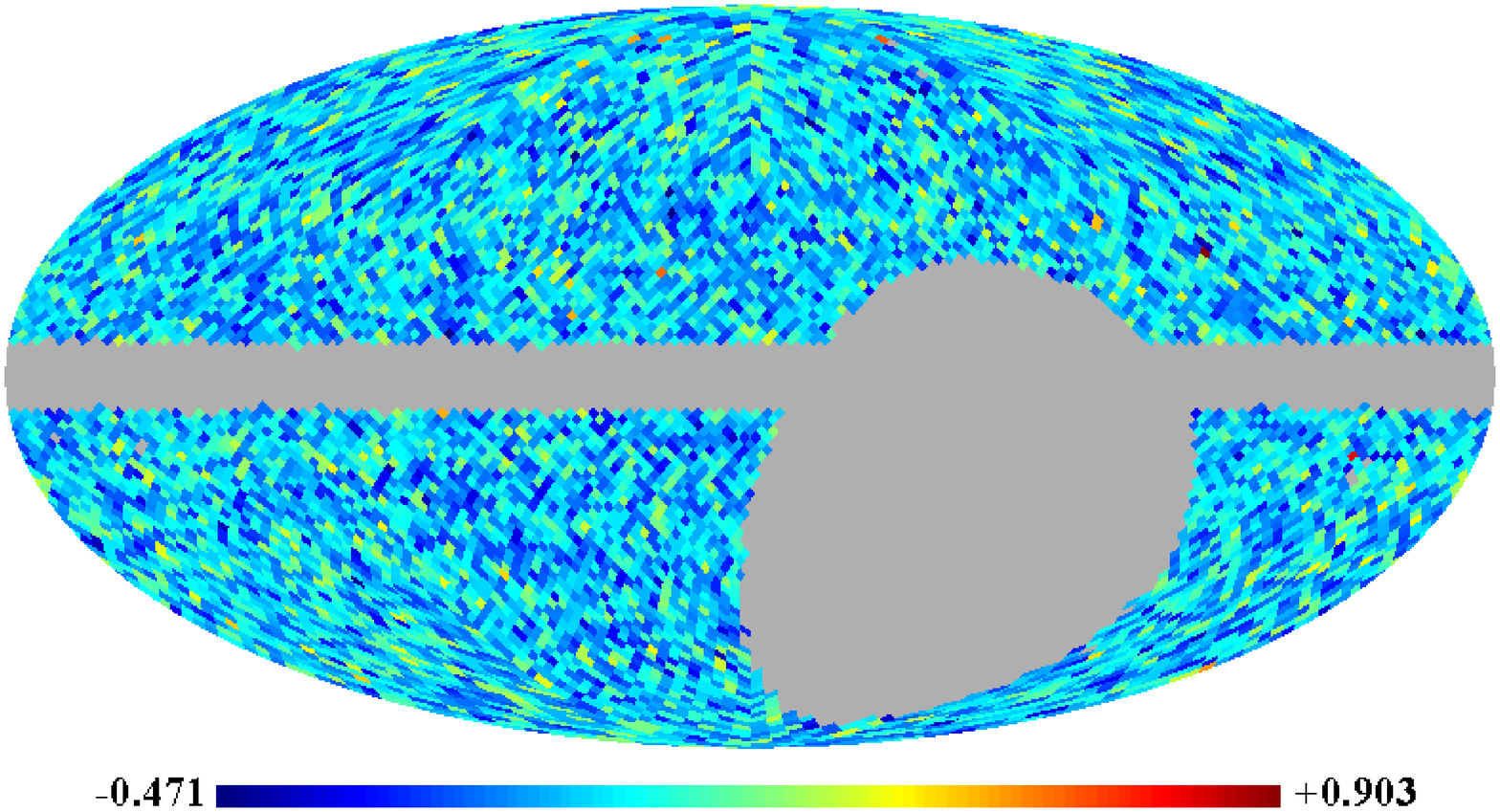}
\caption{\textsc{HEALPix} maps of fluctuations of the NVSS number density of galaxies at $\mathrm{N_{side}}=32$, computed for different flux threholds. 
From top to bottom, the maps correspond to a flux threshold of: $2.5$, $5.0$ and $10.0$ mJy respectively.} 
\label{fig:nvss_data}
\end{figure}

We construct several dimensionless \textsc{HEALPix} \citep{Gorski2005} maps of the fluctuations of the number density of galaxies $\delta(\mathbf{x})$ 
from NVSS. This catalogue is a $1.4 \mathrm{GHz}$ continuum total intensity and linear polarisation survey which explores the largest sky-coverage 
so far (the sky north of J2000.0 $\delta \ge -40^\circ$) and gives reliability in order to consider that the objects which appear are extragalactic. 
Although active galactic nuclei (AGNs) are the dominant contribution in radio catalogues at $1.4$ GHz, \citet{Condon1998}
showed that star-forming galaxies constitute about $30\%$ of the NVSS sources above $1$ mJy. However, this portion
decreases rapidly as higher flux thresholds are considered. The star-forming galaxies of NVSS are nearby sources and they might distort the global pattern
\citep{PlanckXIX2013}. In particular, three different cases are explored in this paper: we only take into account those sources whose flux value is greater 
than $2.5$, $5.0$ and $10.0$ mJy respectively. All these maps are created at a
\textsc{HEALPix} resolution of $\mathrm{N_{side}} = 32$ and are shown in Figure \ref{fig:nvss_data}.  

In addition, it is well-known that the NVSS data present systematic effects due to the adoption of two different configurations, depending on the 
declination angle \citep[e.g.,][]{Blake2002}. We follow the procedure explained by \citet{Marcos2013} in
order to correct the variation of the mean number density of galaxies with the declination angle. The map is divided into $70$ stripes which cover equal
area. Taking into account the mean value in each stripe, the galaxy counting per pixel is rescaled so that the mean value in each band is the same than the mean
computed within the whole map. We only make this correction for the threshold of $2.5$ mJy, because in the rest of cases we assume that
the systematic effects are negligible. We have carried out a series of analyses to check whether this declination correction could either introduce a fake modulation
pattern or, conversely, mitigate a real one. Our simulations show that this is not the case. Even more, we have also analysed the NVSS data without any correction for the
$2.5$ mJy flux cut, and no difference has been found.

The exlusion mask has been constructed as a combination of two requeriments. On the one hand, we impose a threshold in declination 
(excluding pixels with $\delta < -40^{\circ}$), because NVSS only comprises observations performed from the Northern hemisphere and the tropical latitudes.
On the other hand, we exclude a region of $14^{\circ}$ wide which covers the galactic plane in order to discard a possible contamination from 
galactic objects. 

Finally, we consider the estimation of the galaxy power spectrum $C_{\ell}^{\mathrm{GG}}$ proposed by \citet{Marcos2013}. 
Since there are difficulties to describe theoretically the NVSS data due to a power excess presented at large scales, many methods have been used to model the statistical properties of this survey
\citep[e.g.,][]{Dunlop1990, Boughn2002, Ho2008, deZotti2010}. \citet{Marcos2013} performed a joint fitting of the NVSS 
power spectrum and the distribution of galaxies as a function of redshift in the Combined EIS-NVSS Survey Of Radio Sources \citep[CENSORS;][]{Best2003, Brookes2006}.
In particular, they used a gamma distribution in order to parameterize the redshift distribution of galaxies.

\section[Results]{Results}
\label{sec:results_dip}
In this section, we discuss the expected results of the method in the case in which the LSS presents a dipole modulation like that found in the CMB. 
And, then, we also present the application to the NVSS data.

\subsection[Forecast]{Amplitude estimation forecast}
We test the sensitivity of NVSS to detect a dipole modulation as that found in the CMB data, by simulating a NVSS-like realisation. A flux threshold
of $5.0$ mJy, the model proposed by \citet{Marcos2013} and the Poisson noise are considered. We use \textsc{HEALPix} to construct an isotropic Gaussian map at
$\mathrm{N_{side}} = 32$ of the $\boldsymbol{\delta}$ field from the theoretical power spectrum $C_{\ell}^{\mathrm{GG}}$.

We include a multiplicative dipole modulation as that found by \citet{Hoftuft2009}
in the CMB temperature anisotropies, i.e., with an amplitude $A_m = 0.072$ and a preferred direction which points towards $(l,b) = (224^{\circ},-22^{\circ})$. 
Since the propagation of anisotropies, due to the the Sachs-Wolfe and the integrated Sachs-Wolfe effects, is well described by a linear regime, one of 
the possible cases is that the relative amplitude of the dipole modulation is similar to that found in the CMB.

\begin{figure*}
\centering
\includegraphics[scale=0.27]{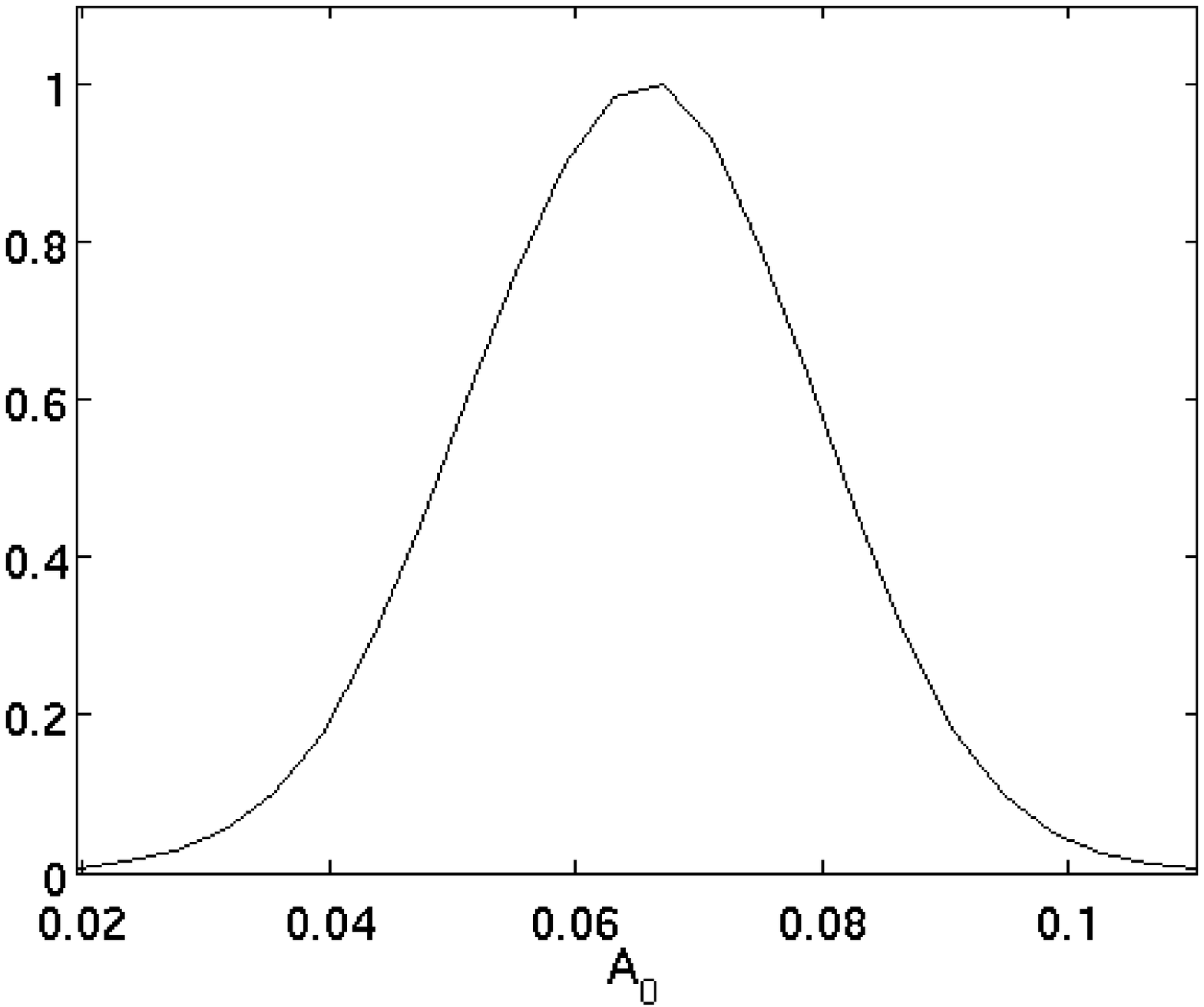}
\includegraphics[scale=0.27]{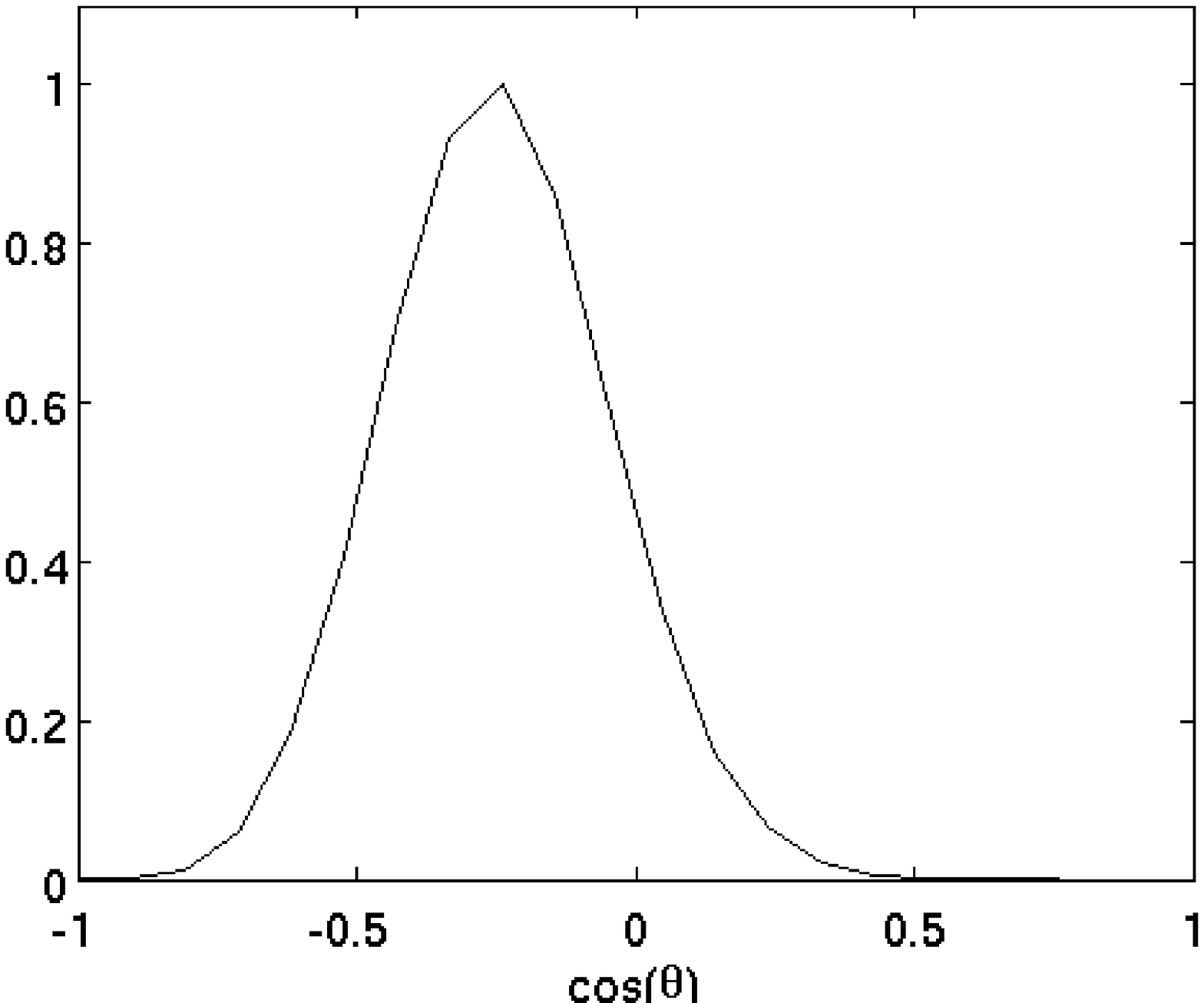}
\includegraphics[scale=0.27]{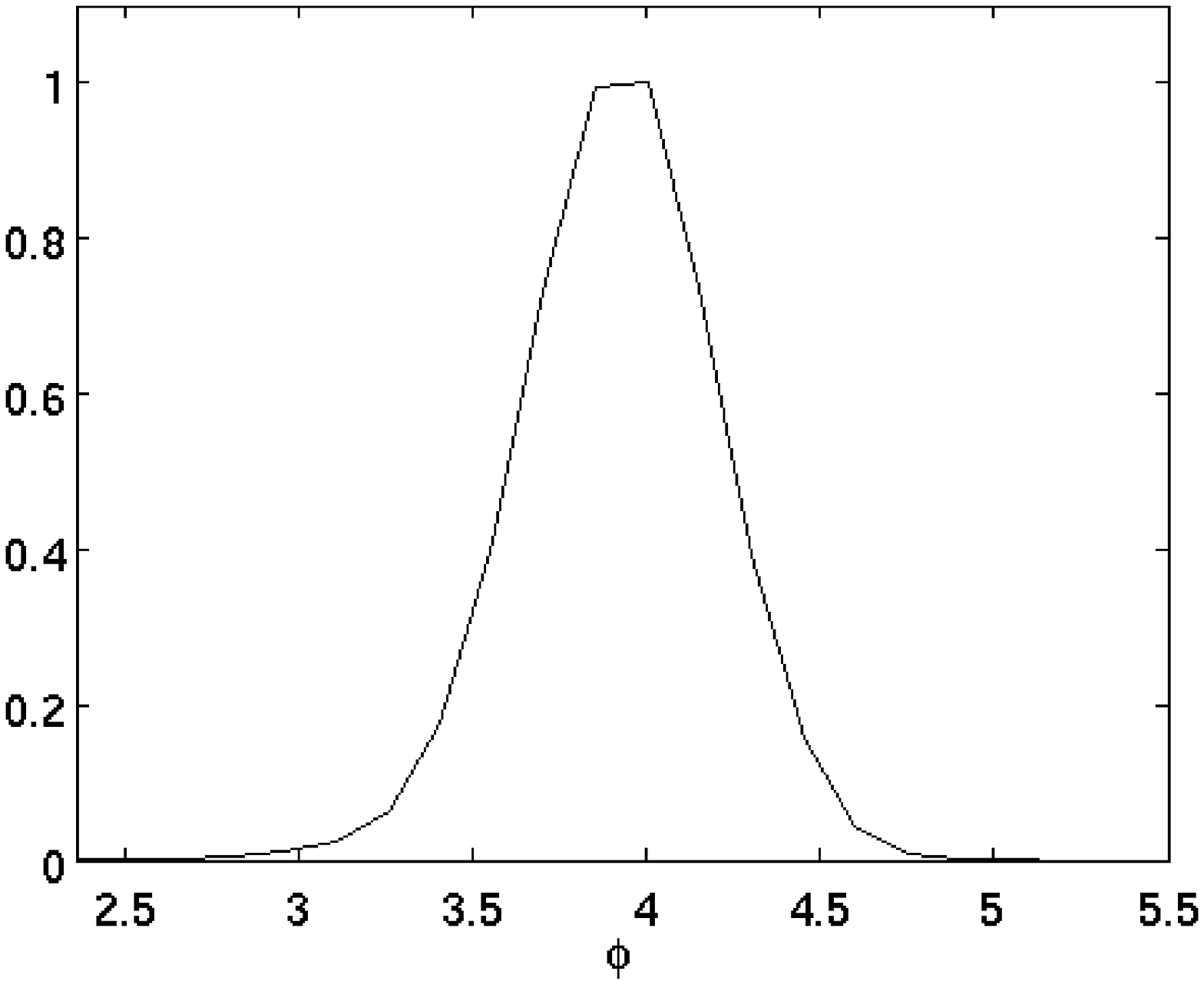}
\caption{Marginalised likelihoods of the parameters of the dipole model for a simulated NVSS map with a flux threshold of $5.0$ mJy
and a dipole modulation with an amplitude $A_m = 0.072$ and a preferred direction which points towards $(l,b) = (224^{\circ},-22^{\circ})$.} 
\label{fig:simulation_est}
\end{figure*}

A detection of a non-negligible dipole modulation 
is obtained with a sensitivity of $5.5\sigma$. The direction of the dipole modulation is recovered with an uncertainty of $17^{\circ}$. The shapes of the marginalised likelihoods for the three parameters of the dipole modulation are shown in Figure \ref{fig:simulation_est}. 

\subsection{Application to the NVSS data}

An estimation of the parameters which characterise a hypothetical dipole
modulation is computed for three different cases, with a flux threshold of 
 $2.5$, $5.0$ and $10.0$ mJy respectively. We conclude that no detection of a dipole modulation is found in any case. The estimated values for 
the amplitude are shown in Table \ref{table:ampl_nvss}. All of them are compatible with zero at a significance level less than $1 \sigma$. 

\begin{table}
\centering
\begin{tabular}{ccc}
\hline
Threshold (mJy) & $\bar{n}_s$ & $A_m$  \\
\hline
$2.5$ & $158594.34$ & $0.003 \pm 0.015$  \\
$5.0$ & $92273.41$ & $0.011 \pm 0.016$  \\
$10.0$ & $55496.39$ & $0.007 \pm 0.014 $ \\
\hline
\end{tabular}
\caption{Marginalised amplitude of a dipole modulation in the NVSS for different flux thresholds. The average number of counts per steradian is denoted by $\bar{n}_s$.}
\label{table:ampl_nvss}
\end{table}

This absence of dipole modulation contrary to the findings in the CMB temperature anisotropies suggests that the dipole
 asymmetry is not caused by a secondary anisotropy located at $z \sim 1$. The result does not support the hypothesis of an anisotropic breaking
 during an early phase of the universe, because the modulation should be linearly propagated to the LSS distribution. Then, a possible cause of the
 dipole modulation found in the CMB could be sought in catalogs of nearby objects, since it is more likely that the local structure 
 is statistically deviated from the homogeneous and isotropic overall pattern. But this is not the only possibility. For instance, there could be 
 topological defects whose distribution generates an hemispherical asymmetry at higher redshift. In particular, the number of textures 
 expected by models is sufficiently small to present an anisotropic distribution \citep{Cruz2007}.

As the mean value of the redshift distribution of NVSS is $\bar{z}=1.2$, more local surveys, such as the Wide-field Infrared Survey Explorer
 \citep[\textit{WISE};][]{Wright2010} may be useful in forthcoming research. In this sense, we also explored the Two Micron All-Sky Survey
 \citep[\textit{2MASS};][]{Skrutskie2006} data, by applying the same methodology that is described in the present paper. However, the difficulty 
 to avoid the very nearby structure prevented us to reach any conclusion. This survey was used by several authors in order to reconstruct the 
integrated Sachs-Wolfe field, which could trace a secondary anisotropy that define a preferred axis on the sky \citep[e.g.,][]{Francis2010, Rassat2013}.


\section{Conclusions}
\label{sec:Conclusions}

After the confirmation of the CMB statistical anomalies by the \textit{Planck} results \citep{PlanckXXIII2013}, the LSS provides an alternative observable
to study the origin of these deviations. In particular, an asymmetry usually parameterized as a dipole modulation across the sky was detected
by several authors in the CMB temperature anisotropies. We adapt here the method described by \citet{Eriksen2007} to deal with LSS data.

Two possibilities are considered: the asymmetry could be due to an intrinsic isotropy breaking occurred in
the early Universe or it could be caused by a secondary anisotropy induced by an anisotropic distribution of the local galaxy
distribution. If different LSS surveys presented a sort of dipole modulation with a similar preferred direction than that observed
in the CMB temperature anisotropies, it would be an indicator that the cause of this anomaly has to be sought in the physical
mechanisms of generation of primordial fluctuations. But, if this preferred direction was not detected at all in LSS data,
the CMB asymmetry could be due to, for instance, a secondary anisotropy located in a more local or further galaxy distribution 
than the one traced by the surveys we are considering.

The methodology is proven reliable with a NVSS-like simulation with a dipole amplitude as intense as that measured in the CMB data. However, no preferred direction
is detected in the NVSS data for three different flux thresholds: $2.5$, $5.0$ and $10.0$ mJy respectively. 

Assuming a linear propagation of the dipole modulation hypothetically generated during an early phase of the universe, the apparent absence of detection in the LSS at $z \sim 1$ 
suggests that the dipole power modulation found in the CMB had to be generated in a late cosmological epoch different from this one.
Avoiding other considerations, those models based on anisotropic modifications of standard inflation \citep[e.g.,][]{Donoghue2009, McDonald2013, Damico2013} might be compromised to reconcile their predictions with observation. This result motivates forthcoming studies with other surveys which explore the more local structure distribution. A detection
of a dipole modulation in the nearby galaxy distribution might explain the CMB asymmetry in terms of a secondary anisotropy. 


\section*{acknowledgments}
We acknowledge partial financial support from the Spanish \textit{Ministerio de Econom\'ia y Competitividad} Projects HI2008-0129, AYA2010-21766-C03-01, AYA2012-39475-C02-01 and Consolider-Ingenio 2010 CSD2010-00064. 
RFC thanks financial support from Spanish CSIC for a \textit{JAE-predoc} fellowship, co-financed by the European Social Fund. The authors acknowledge 
the computer resources, technical expertise and assistance provided by the \textit{Spanish Supercomputing Network} (RES) node at Universidad de Cantabria. 
We acknowledge the use of the NASA's HEASARC archive. The HEALPix package \citep{Gorski2005} and the \textsc{CosmoMC} code \citep{Lewis2002} were used throughout the data analysis.
\bibliographystyle{mn2e}
\bibliography{citas_dipole}

\end{document}